\documentclass[acmlarge]{acmart}

\makeatletter
\newcommand{\myconfshort}{\acmConference@shortname}
\newcommand{\myconffull}{\acmConference@name}
\newcommand{\myconfdate}{\acmConference@date}
\newcommand{\myconfloc}{\acmConference@venue}

\AtBeginDocument{
  \fancypagestyle{firstpagestyle}{
    \fancyhead{}%
    \fancyfoot[C]{}%
  }
  \fancyhf{}
  \fancyhead[LO]{\@headfootfont\shorttitle}%
  \fancyhead[RE]{\@headfootfont\@shortauthors}%
  \fancyhead[LE]{\@headfootfont\footnotesize \myconfshort, \myconfdate, \myconfloc}%
  \fancyhead[RO]{\@headfootfont\footnotesize \myconfshort, \myconfdate, \myconfloc}%
  \fancyfoot[C]{}%
}
\makeatother

\acmBooktitle{\conffull\@ (\confshort), \confdate, \confloc}

\usepackage{array}
\usepackage{enumitem}
\usepackage[title]{appendix}

\copyrightyear{2026}
\acmYear{2026}
\setcopyright{cc}
\setcctype{by}
\acmConference[FAccT '26]{The 2026 ACM Conference on Fairness, Accountability, and Transparency}{June 25--28, 2026}{Montreal, QC, Canada}
\acmBooktitle{The 2026 ACM Conference on Fairness, Accountability, and Transparency (FAccT '26), June 25--28, 2026, Montreal, QC, Canada}
\acmDOI{10.1145/3805689.3806719}
\acmISBN{979-8-4007-2596-8/2026/06}

\begin{document}

\title[Reimagining Open Source and Openness in AI]{Reimagining Open Source and Openness in AI: Co-Creating Responsible Technological Futures}

\author{Genevieve Smith}
\email{genevieve.smith@stanford.edu}
\affiliation{%
  \institution{Stanford University}
  \city{Palo Alto}
  \country{USA}
}

\author{Hiral Patel}
\affiliation{\institution{UC Berkeley}\city{Berkeley}\country{USA}}

\author{Steven Luo}
\affiliation{\institution{UC Berkeley}\city{Berkeley}\country{USA}}

\author{Monica G. Bobra}
\affiliation{\institution{State of California}\city{Sacramento}\country{USA}}

\author{Judy Brewer}
\affiliation{\institution{UC Berkeley}\city{Berkeley}\country{USA}}

\author{Cathryn Carson}
\affiliation{\institution{UC Berkeley}\city{Berkeley}\country{USA}}

\author{Isadora Cruxen}
\affiliation{\institution{Queen Mary University of London}\city{London}\country{UK}}

\author{Shachee Doshi}
\affiliation{\institution{Independent}\city{Washington D.C.}\country{USA}}

\author{Maximilian Gahntz}
\affiliation{\institution{Mozilla}\city{Munich}\country{Germany}}

\author{Nicholas Garcia}
\affiliation{\institution{Independent}\city{San Francisco}\country{USA}}

\author{Natalia Luka}
\affiliation{\institution{UC Berkeley}\city{Berkeley}\country{USA}}

\author{Meredith M. Lee}
\affiliation{\institution{UC Berkeley}\city{Berkeley}\country{USA}}

\author{Min Kyung Lee}
\affiliation{\institution{UT Austin}\city{Austin}\country{USA}}

\author{Woohyeuk Lee}
\affiliation{\institution{UT Austin}\city{Austin}\country{USA}}

\author{Jarrod Millman}
\affiliation{\institution{UC Berkeley}\city{Berkeley}\country{USA}}

\author{Ricardo Miron Torres}
\affiliation{\institution{Digital Public Goods Alliance}\city{León}\country{Mexico}}

\author{Chinasa T. Okolo}
\affiliation{\institution{Brookings Institution}\city{Washington D.C.}\country{USA}}

\author{Cailean Osborne}
\affiliation{\institution{University of Oxford}\city{Oxford}\country{UK}}

\author{Derek Slater}
\affiliation{\institution{Independent}\city{San Francisco}\country{USA}}

\author{Katie Steen-James}
\affiliation{\institution{Open Source Initiative}\city{San Francisco}\country{USA}}

\author{Nikko Stevens}
\affiliation{\institution{Massachusetts Institute of Technology}\city{Cambridge}\country{USA}}

\author{Jennifer Tridgell}
\affiliation{\institution{University of Cambridge}\city{Cambridge}\country{UK}}

\author{David Gray Widder}
\affiliation{\institution{UT Austin}\city{Austin}\country{USA}}

\renewcommand{\shortauthors}{Smith et al.}

\acmArticleType{Research}

\acmContributions{}

\begin{abstract}
\section*{Abstract}
Debates over open source and openness in artificial intelligence (AI) have intensified as policymakers, researchers, and practitioners grapple with how foundation models should be developed and governed to balance innovation, accountability, and public interest. However, there has been limited empirical work examining how diverse stakeholders collectively understand and negotiate responsible openness in AI, particularly through participatory processes that extend beyond industry-led definitions and frameworks. This paper presents findings from a multi-sectoral workshop grounded in futures thinking and participatory design methods. The workshop generated co-created visions of desirable futures and the role of AI, alongside a set of action pathways and a research roadmap focused on responsible open source and openness in AI. This paper makes three key contributions. First, it empirically documents the co-created visions, actions, and research priorities. Second, it identifies four core tensions that emerged as participants translated high-level aspirations into concrete actions, revealing conflicting interpretations of openness regarding its purpose (as an end or a means), its scope (expansion versus meaningful access), and its operation (mandatory versus conditional, sufficient versus dependent on governance and use). These tensions illustrate that responsible openness is not a singular technical solution, but a negotiated sociotechnical project shaped by values, positionalities, and priorities. Third, the paper advances methodological approaches in AI governance by demonstrating how participatory futures methods can surface plural visions, actions, and research priorities that extend beyond dominant, largely corporate, narratives. This work contributes to FAccT’s growing sociotechnical and participatory turn by bringing futures thinking and participatory design methods to AI governance and agenda-setting around open source AI, while offering empirical insight into how openness, power, and accountability are negotiated in practice.\footnote{This is the author's version of the work. This preprint is posted here for your personal use. The definitive Version of Record is published in the 2026 ACM Conference on Fairness, Accountability, and Transparency (FAccT '26), June 25--28, 2026, Montreal, QC, Canada, http://doi.org/10.1145/3805689.3806719.}
\end{abstract}

\vfill
\enlargethispage{2\baselineskip} 
\vspace*{-1.5pc}                 

\maketitle

\section{Introduction}
\label{sec:introduction}
The future of AI is at a critical juncture, with growing debate over whether---and to what extent---foundation models should be open, and how their development and governance can minimize harm while realizing public benefit. Open source and open AI models promise to advance research, innovation, transparency, and equitable participation \cite{eiras2024} \cite{seger2023}. Historically, open source has helped diffuse innovation, challenge incumbents, and enable viable alternatives in otherwise concentrated markets \cite{weber2004success}. In the context of generative AI, openness has taken on new meanings and stakes. Openness alone does not guarantee accountability, equity, or democratic control, while models labeled as “open” can reproduce structural inequities or obscure critical aspects of their creation and governance \cite{widder2024nature}. Moreover, whether models are open or closed, key design decisions---such as data sources, training processes, and access conditions---are often shaped by the values and market priorities of a narrow set of powerful actors, particularly large technology firms \cite{widder2024nature}. These dynamics point to a deeper issue: definitions and practices of openness and responsibility in AI are not neutral technical classifications, but sites of agenda-setting power. What counts as “open”, who decides where systems fall along the openness spectrum, and who has the capacity to meaningfully engage with them shape which futures become imaginable, legitimate, and actionable. In response, we conducted a workshop grounded in participatory futures and speculative design with diverse, multidisciplinary, and multi-sectoral leaders to collectively imagine responsible openness in AI and produce co-created visions alongside concrete action pathways and a research roadmap.

This paper makes three contributions. First, it empirically documents the co-created visions, action pathways, and research priorities to advance responsible openness in AI that resulted from the workshop. Second, it identifies four tensions that emerged as participants translated high-level aspirations into actions, revealing conflicting interpretations of openness regarding its purpose (as an end or a means), its scope (expansion versus meaningful access), and its operation (mandatory versus conditional, sufficient versus dependent on governance and use). These tensions illustrate that responsible openness is not a singular technical solution, but a negotiated sociotechnical project shaped by values, positionalities, and priorities. Third, the paper advances methodological approaches in AI governance by demonstrating how participatory futures methods can surface plural visions, actions, and research priorities that extend beyond dominant, largely corporate, narratives. This work contributes to FAccT’s growing sociotechnical and participatory turn by bringing futures thinking and participatory design methods to AI governance and agenda-setting around open source AI, while offering empirical insight into how openness, power, and accountability are negotiated in practice. 

This paper proceeds as follows. The background section (\ref{sec:related-work}) explores understandings and practices of openness and open source in AI, situating these debates within broader questions of power and accountability. We detail the methodology (\ref{sec:methodology}) before turning to our findings (\ref{sec:findings}), followed by a discussion (\ref{sec:discussion}) and a conclusion (\ref{sec:conclusion}).   

\section{Background}
\label{sec:related-work}
In this background section, we outline the meanings and practices of open source and openness in AI, as well as consider the risks and benefits. We then turn to how power operates in the landscape of open and open source AI before discussing emerging movements that seek to envision alternative, participatory technological futures. 

\subsection{What is open source and openness in AI?}
Openness has long been central to the ethos and practice of software development, yet in the context of generative AI it has taken on new meanings and stakes \cite{Paris2025OpenAI}. Scholars warn of “open-washing”, whereby actors market systems as “open” to influence policymakers, regulators, and consumers, while withholding key components needed for meaningful scrutiny or reuse \cite{Liesenfeld2024}. As a result, debates over openness in AI increasingly center not only on access to models, but on which elements of the AI stack are disclosed, to whom, and under what conditions. While there is no single agreed-upon definition of open or open source AI (see Appendix \ref{sec:appendix} for commonly referenced terms), the Open Source Initiative’s Open Source AI Definition (OSAID) has emerged as an influential reference point, defining open source AI as systems that allow users to (1) use the system for any purpose without needing permission, (2) study how the system works and inspect its components, (3) modify the system for any purpose, including to change its output, and (4) share the system for others to use with or without modifications, for any purpose \cite{OSAID}. Notably, this definition does not mandate releasing full training data, instead allowing “sufficiently detailed information” about it. This has foregrounded ongoing debates about what should count as “source” in AI systems and where openness should be bounded. 

In practice, openness in AI exists on a spectrum shaped by developer decisions about which components of the AI stack---such as documentation, model weights, source code, or training data---are made public \cite{eiras2024, solaiman2023gradientgenerativeairelease}. Open or open source AI models are frequently equated with open weights, yet weights alone do not ensure full transparency, reproducibility, or accountability when other critical components remain closed. High-profile models illustrate this variability. For example, Meta released LLaMA-2 with weights and architecture but is not recognized as open source by the European Union or OSI \cite{OSI2025MetaLlama}. DeepSeek R1, created by Chinese developers, similarly released weights and training methods without disclosing training data, leading it to be classified by the Model Openness Framework \cite{white2024} as a Class III “Open Model” \cite{LinuxFoundation2025IsItOpen} in January 2025. 

Academic and nonprofit actors have advanced models more closely aligned with the OSAID, offering full transparency and access to components such as model architecture, weights, and documentation. For example, researchers at the MIT Jameel Clinic for Machine Learning in Health introduced Boltz-1, the first fully open source model (including training data) that achieves state-of-the-art performance \cite{zewe2024mit}. The Allen Institute for AI (Ai2) has released powerful, fully open source models and evaluation frameworks \cite{AllenAIModels, Bishop2025}. In addition, much of the early ideation process happens on blogs and online communities such as The Gradient, LessWrong, and X. Despite technical contributions, such projects typically receive less visibility and use than large corporate models, highlighting asymmetries in attention and agenda-setting in the field. 

The open AI ecosystem is expanding rapidly. Model parameters on Hugging Face increased 17-fold from 2020 to 2025, with a general upward trend towards multimodal and computationally efficient architectures \cite{longpre2025economies}. There has also been growth in intermediary organizations that specialize in re-packaging models through fine-tuning, quantizing, or building artistic adapters \cite{longpre2025economies}. Prior work shows that collaboration around open LLMs extends to datasets, benchmarks, open source frameworks, leaderboards, and compute partnerships \cite{linaker2025cartography}. In 2023, individual contributors to generative AI projects grew by 148\% and developers made 301 million contributions to open projects across GitHub, including widely used projects such as Stable Diffusion and LangChain \cite{Daigle2024GitHub}. While questions remain around sustainability and modes of collaboration across model lifecycles \cite{choksi2025brief}, the scale and pace indicate that open AI is a durable part of the AI landscape. 

\subsection{Risks and benefits of open source and openness in AI}
\label{sec:Risks}
Open source and openness in AI present both significant promise and risk. Open source models offer immense promise for societal impact compared with closed source models across key areas such as research, innovation, inclusivity, and transparency; enabling those with technical proficiency to quickly test ideas, experiment with new models, and iterate on existing work \cite{seger2023}. Open source tools lower barriers to entry for new developers and researchers, and encourage a culture of sharing, collaboration and learning, helping to build a more inclusive and knowledgeable AI community; while also allowing for outside scrutiny and auditing \cite{langenkamp2022}. Adoption of open models is already widespread among enterprises \cite{finley2025github}, and governments increasingly frame openness as a strategic priority, including in the White House’s 2025 America’s AI Action Plan \cite{whitehouse2025aiplan}. Meanwhile, the French government invested in the core developer team of scikit-learn, a Python library for ML with over 3.5 billion downloads, to build an open source commons for data science and AI via its 2021 national AI strategy as a means to advance national competitiveness in AI, AI adoption by small and medium-sized enterprises, as well as digital sovereignty \cite{osborne2024}. On the flip side, open AI models can introduce risks, particularly given their ability to be modified, repurposed, and redistributed without centralized control \cite{hendrycks2023overview, seger2023}. 
However, recent work emphasizes that many harms stem less from openness than from availability and capability, with widely available proprietary models also enabling misuse at scale \cite{casper2025open}. Evaluating risks therefore requires attention not only to absolute risk, but the \textit{marginal risk} introduced by openness relative to closed alternatives and existing technologies, as well as to gaps in current governance and accountability mechanisms for open systems \cite{bommasani2023}. 

At a higher level, current AI development risks reinforcing monopolization and consolidating power in a handful of dominant tech companies \cite{furlong2024}. As models grow in size and capability, the costs of training, deployment, and maintenance increasingly privilege actors with access to large-scale compute, data, and financial resources. These advantages are further amplified by massive infrastructural investments, like data centers or processor chips, that enable vertically integrated ecosystems that are difficult for others to replicate. Restricted access to closed models accessed through APIs intensifies this consolidation. Meanwhile, selective openness may strategically serve to strengthen technical standards and ecosystems and steer innovation, without meaningfully redistributing control \cite{widder2024nature}. As the history of open source shows, such practices can allow dominant firms to entrench their market position while extracting value from community contributions \cite{widder2024nature}. As such, the choice between open and closed is often less a binary and more a strategic calculation---shaped by business models, regulatory pressures, and long-term positioning. 

As this field evolves, so do the global stakes. Open AI systems are increasingly framed as tools for national capacity building and digital public goods \cite{adams2025}, aligning with multilateral initiatives that promote open technologies for public benefit (e.g., the United Nations' Digital Public Goods and the Digital Public Goods Alliance). Open AI systems offer practical advantages by enabling reuse, adaptation, and fine-tuning of existing tools that can help mitigate the consequences of persistent asymmetries in access to compute and infrastructure that remains consolidated in the Global North. Governments and international organizations are responding with a range of governance initiatives---binding legal frameworks, regulations, harmonized regional strategies, and high-level ethical statements---that seek to leverage openness while mitigating associated risks.\footnote{See for example, the United Nations’ Global Dialogue on AI Governance and Independent International Scientific Panel on AI, the OECD’s updated AI Principles, the Council of Europe’s Framework Convention on Artificial Intelligence, the African Union’s Continental AI Strategy, the ASEAN Guide on AI Governance and Ethics, the European Union’s Regulation (EU) 2024/1689 (AI Act), the United States’ Executive Order on Ensuring a National Policy Framework for Artificial Intelligence, China’s Generative AI Regulation, and Brazil’s Bill 2338/2023.} At the same time, these systems are not neutral, as even openly available models embed particular priorities and perspectives shaped by their development contexts. 

\subsection{Concentration of agenda-setting power in open AI}
\label{sec:concentration}
Openness in AI is shaped by multiple, overlapping forms of power. The first is \textit{definitional power}, or the authority to determine what "openness" itself means and whose interpretations are legitimized. As Indigenous, feminist, and postcolonial scholars remind us, designations of "openness" are never purely technical; rather they are acts of boundary-making that delineate whose participation and epistemologies count \cite{kukutai2016indigenous, chan2019open, carroll2021operationalizing}. The current AI field is shaped by rapid cycles of innovation and hype, with dominant generative AI narratives and agendas largely led by a narrow set of powerful actors (particularly large technology firms, but also influential standards bodies and platform owners) who exercise disproportionate influence in how openness is defined, operationalized, and legitimized. This concentration of agenda-setting power and dominance of a small set of corporate actors risks narrowing the imaginaries of AI futures, privileging commercial interests over broader social concerns \cite{benjamin2019, crawford2021, birhane2021algorithmic}. Further, the concentration of agenda-setting power influences which problems, risks, and governance approaches are prioritized in policy and research. 
Recent work has shown that definitions of AI openness are not neutral classifications, but governance tools shaping regulation, liability, and institutional incentives \cite{Paris2025OpenAI}. Openness is often defined narrowly in terms of afforded actions at the model level, while broader effects are assumed rather than ensured, obscuring how openness operates at multiple scales \cite{Paris2025OpenAI}. 

The second and related form of power is \textit{infrastructural power}, which is control over what is actually made open, when, and under what conditions. While openness is often celebrated as central to transparency and democratization, scholars have warned that claims of openness can amount to “open washing”, where selective disclosure results in reputational benefits without meaningful reproducibility, accountability or redistribution of power \cite{Liesenfeld2024}. Training data has emerged as a particularly salient site of contestation. Recent efforts to define “open source AI” (e.g., the OSAID), have foregrounded longstanding debates about whether and when training data should be considered part of an AI system’s “source”. Some argue data is functionally analogous to source code, as model weights encode statistical properties derived from it, while others point to legal, ethical, and practical constraints (e.g., privacy, licensing, consent) that complicate full release. These disagreements reflect broader tensions between reproducibility, accountability, and protection from harm, and remain unresolved. 
Infrastructural power extends beyond data, including compute, platforms, and development ecosystems. Prior work demonstrates that control over large-scale compute, toolings, and development frameworks shapes who can build, evaluate, and meaningfully participate in AI systems \cite{widder2024nature}. Even when code or models are open, limited access to compute constrains participation and concentrates influence among a small set of actors. Platforms such as PyTorch further illustrate how infrastructural control translates into agenda-setting power by enabling owners to shape default architectures, workflows, and benchmarks. In doing so, technical and research directions of the broader field are shaped. 

A third and often overlooked form of power is \textit{normative or value-encoding power}, which is the capacity to embed particular ethical frameworks and values into how open or open source models are developed and managed. Software licenses have long been understood not merely as legal instruments, but as expressions of the values their authors seek to uphold \cite{choksi2024opensource}. Even in open or partially open systems, authority over design, release, and governance decisions remains concentrated among creators and other contributors, who effectively determine which values are built into the system through training data, alignment processes, and usage constraints. For example, models trained or deployed within specific institutional or national contexts may exclude or suppress categories of information deemed politically or culturally sensitive, reflecting the particular agendas or ideological frameworks of their developers. This has been illustrated with DeepSeek, which users found does not talk about Tiananmen Square \cite{lee2025bloomberg} and reflects “Chinese government talking points” \cite{politico2025deepseek}. Ongoing debates around "responsible" open source therefore hinge not only on practices of release and transparency, but on who gets to decide what counts as responsible, whose values those definitions reflect, and whose interests they serve. International bodies, such as UN Conference on Trade and Development, have warned that without global frameworks to enforce transparency and accountability, these value choices risk reinforcing existing power asymmetries rather than advancing AI towards public interest \cite{unctad2025technology}. In this sense, value-encoding power becomes a subtle yet persistent form of agenda setting, as those who define what constitutes harm, risk, or acceptability effectively determine the boundaries of legitimate inquiry and debate; thereby informing not only how AI operates, but which social priorities and perspectives are elevated or marginalized. 

Taken together, these dynamics illustrate that definitions of openness are not merely technical, but sites of agenda-setting power. What counts as “open”, who decides where systems fall along the openness spectrum, and who has the capacity to meaningfully engage with them all shape whether openness redistributes or entrenches power. It is this concentration of definitional, infrastructural, and normative power that motivates this research, including the importance of incorporating diverse, multidisciplinary voices in the shaping of AI systems \cite{costanzachock2020, bender2021}. 

\subsection{Envisioning alternative technological futures}
An established body of work has long called for envisioning alternative technological futures, particularly those grounded in diverse social and cultural perspectives \cite{haraway1988situated, harding1991, dunne2013, costanzachock2020}. Building from feminist science and technology studies (STS), design justice, and speculative design traditions, this scholarship emphasizes that technology is not neutral, but rather shaped by the values, experiences, knowledge, and positionalities of its developers. These calls extend to AI, where scholars and practitioners seek to surface alternative imaginaries that foreground inclusion, equity, and justice \cite{bender2021, benjamin2024}. Participatory and co-design methods have been central to this effort. These approaches move beyond technical abstraction to support collective imagination and critique. For example, recent work convened diverse participants to co-create speculative artifacts (i.e. board games) as a means of interrogating sociotechnical questions around technology impacts and to identify desirable technological futures \cite{odendaal2024}. Within the FAccT community there has been a growing engagement with participatory design and co-design approaches to AI governance and development \cite{scott2022, tseng2025, suresh2022}, as well as interventions to advance stakeholder involvement for responsible AI \cite{kallina2025stakeholder}. While explicit futures thinking remains nascent in FAccT core proceedings, CRAFT sessions have engaged in speculative design and futures thinking (see, for example \cite{vandenboom2025}). This research builds on such scholarship and work.

Futures-oriented approaches extend participatory design by inviting communities to imagine and deliberate on alternative technological futures. Methods of Futures Workshops \cite{vidal2005} and Participatory Speculative Design draw on structured facilitation of collective imagination and critique to support visioning and pathway building. Futures Workshop methods have been applied in various domains as a tool for collective problem solving, including organizational strategy and public-sector planning \cite{terranova2008}. Other researchers integrated the Afrofuturist Speculative Design Toolkit \cite{bray2022} to envision technological and social utopias and dystopias with Black youth in South Side Chicago \cite{harrington2021}. These methods foreground plurality and productive tension rather than consensus, highlighting the role of “contamination”, the process through which distinct entities become entangled, in collective world-building \cite{lu2024}. 

Recent initiatives, including the Columbia Convening on Openness in AI \cite{bdeir2024introducing}, explored openness through multi-stakeholder dialogue. Our workshop extends this inquiry and builds on the lineage of participatory futures design to explore how alternative futures of responsible openness might be collectively imagined, negotiated, and enacted. 

\section{Methodology}
\label{sec:methodology}
\subsection{Methodological approach: Futures thinking and participatory design}
Through participatory and futures-oriented design methods, we sought to create a setting where diverse publics could co-develop visions of responsible openness in AI.  
The workshop employed a mixed theoretical and participatory approach grounded in feminist STS, speculative design, and participatory design. Drawing on the concept of \textit{}situated knowledges \cite{haraway1988situated}, our methods sought to surface plural perspectives on AI and technological futures recognizing technology design as inherently political and value-laden. This theoretical foundation informed our use of speculative and participatory design traditions, including future workshops \cite{jungk1987future} and other participatory approaches, to imagine inclusive technological futures \cite{bodker2015thirdwave, costanzachock2020}. Collectively, the workshop structure sought to support collective imagination, reflexivity, iteration, and inclusivity in the co-production of knowledge about AI futures.

After an initial day of presentations and panel discussions, co-creation activities occurred on the second day across three participatory sessions employing iterative cycles of brainstorming, sticky-note clustering, theme development, and collective reflection. Session 1 \textit{(Visioning)} focused on envisioning desirable futures ten years ahead, articulating the values that defined them and the role of AI within those visions. Session 2 \textit{(Action planning)} translated these visions into concrete action pathways across eight dimensions. Session 3 \textit{(Research roadmap)} identified open questions, knowledge gaps, and methodological needs, producing a community-driven research roadmap complementing the action plan. 

Participants worked in pre-assigned breakout groups of seven to nine people to ensure diversity across affiliation, sector, and demographic background. The same breakout groups remained together across the three sessions to support continuity and cumulative sensemaking. Each breakout group worked with shared materials, including markers, sticky notes, and flip chart paper; one breakout group included virtual participants and used a pre-configured virtual whiteboard and digital sticky-note space. The workshop was intentionally designed not to pursue consensus, as consensus-oriented deliberation is known to suppress expression of conflicting interests and minority perspectives \cite{mansbridge2006, mouffe2005}. Instead, the workshop sought to surface and challenge diverse perspectives, thereby “opening up” deliberation to support collective problem framing and prevent premature closure around singular solutions \cite{stirling2008}. Through this process of engagement and “contamination,” \cite{callon2009} the design surfaced both common ground and tensions. 

\subsection{Participant selection and recruitment}
Participants were recruited using purposive and snowball sampling approaches. The goal was to create a balance of different forms of knowledge and practice that enabled deep technical or organizational expertise, as well as critical perspectives on social and ethical considerations, with representation across government, nonprofit and civil society organizations, industry, and academia. Invitees included participants with expertise and lived experience in open source software (OSS) and open AI initiatives, as well as AI and equity, human rights, justice, and/or sustainability. An interdisciplinary steering committee informed the initial invitation list. To avoid over-reliance on established networks, participants were encouraged to nominate additional participants through a nomination form. Seventy-five participants registered for the workshop, from a pool of 88 individuals on the invitation list and 12 people identified through the nomination form. While most participants were based in the US, there was global representation from people based in Brazil, France, Germany, Mexico, Nigeria, and the United Kingdom. Participants' primary affiliations included academia (54.67\%), industry (14.67\%), civil society or nonprofit institutions (20\%), government (9.33\%), and foundations (1.33\%). National Science Foundation (NSF) funding enabled travel support for approximately one-third of participants.

\subsection{Data and thematic analysis}
Workshop documentation included summarized photographed artifacts (e.g., clustered post-its and themes), as well as facilitator and group discussion notes. These materials were anonymized, compiled into a qualitative corpus, and analyzed. We conducted inductive thematic analysis \cite{braun2006, charmaz2006} to identify recurring patterns, points of convergence, and tensions. Initial open coding was conducted followed by iterative rounds of discussion to refine categories, group them into higher-level themes, and resolve discrepancies. Tensions were primarily identified through the co-occurrence of difficult-to-reconcile or contradictory proposals as participants translated shared aspirations into concrete actions, rather than through explicit disagreement. Primary coding and synthesis were led by a researcher who also played a lead role in the workshop’s facilitation and steering committee. Interpretation was informed by this researcher’s scholarly background on open source, responsible AI, and sociotechnical questions of power and fairness, while iterative discussions with co-authors, who also participated in the workshop, supported reflexivity and rigor in interpretation. 

\section{Findings}
\label{sec:findings}
The findings trace a progression from shared aspirations to concrete proposals and unresolved tensions. Participants articulated a common vision of desirable futures and the role of AI in supporting them, followed by proposed actions and research directions. While many proposed actions were complementary or could coexist, others surfaced tensions. The sections that follow present this progression in three stages: first, the co-created visions; second, the action pathways and research dimensions; and third, an analytic synthesis of emergent tensions.

\subsection{Co-created visions for technology futures}
\label{sec:visions}
The workshop invited participants to “Imagine it’s ten years from now: picture the world you want to live in.” Participants were asked to describe what that future would look and feel like, what values would define it, and what role AI might play in bringing it about; using backcasting to reason from those futures toward the forms of openness, governance, and design that could make them possible. Participants envisioned a world of shared prosperity, whereby all people have the resources and opportunities to live dignified, fulfilling lives. Central to this vision was a commitment to social and economic justice, including respect for human rights, redistribution of power, and conditions for collective wellbeing. AI was imagined as a tool deliberately developed and governed to advance these goals. 

Across visioning breakout groups were a set of recurring themes that characterized the futures desired and the role of AI within them: social and economic justice, community and human connection, democratic governance, agency, and environmental sustainability. Social and economic justice was often articulated in terms of expanding equality of opportunity, with AI as a means to enable equitable outcomes, democratize access to resources, support human rights, and redistribute power. Community and human connection emphasized designing AI systems that strengthen social bonds and collective resilience rather than replacing or distorting human relationships. Democratic governance was framed around decentralizing power in AI development and oversight, with participatory stewardship and trusted public institutions acting in the public interest. Agency was frequently associated with meaningful consent and personal autonomy in data use and digital interaction; including being informed about how, when, and where their data is used and having the ability to control one’s data on their own terms. Finally, sustainability was invoked in relation to green AI supply chains and aligning technological progress with planetary well-being.

Participants identified openness and transparency as foundational conditions for realizing and sustaining such a world. Openness was framed not simply as a technical attribute of model release, but as a form of shared stewardship in which AI systems are treated as digital commons for collective benefit. Participants emphasized that meaningful collaboration requires openness across the AI stack (i.e. data, models, software, and infrastructure) combined with interoperability and mechanisms for collective input. Transparency was framed less as one-way disclosure, and more as reciprocal accountability among  developers, institutions, and the public. 

\subsection{Action pathways and research roadmap}
\label{sec:actions}
To operationalize the co-created vision, the workshop generated concrete action pathways and a research roadmap toward responsible openness in AI across eight cross-cutting dimensions (see Table 1). Together, these outputs form a layered agenda anchored by a foundational north star (co-created vision), interventions to get there (actions), and extended through future-oriented inquiry (research roadmap). These outputs synthesize proposals from group brainstorming and discussion and do not necessarily reflect the views of individual participants or organizations. While participants broadly converged on shared aspirations during the visioning exercise, greater divergence emerged as they considered how openness should function in practice and what should be prioritized. Many proposed actions and research directions were complementary, while others surfaced tensions where approaches could not easily coexist. This section explores each of the dimensions, with tensions further discussed in the following section. 

\begin{table*}[t]
\centering
\caption{Actions and research priorities toward the co-created vision}
\footnotesize
\setlength{\tabcolsep}{3pt}
\renewcommand{\arraystretch}{1.15}
\begin{tabular}{
  >{\raggedright\arraybackslash}p{3.1cm}
  >{\raggedright\arraybackslash}p{4.0cm}
  >{\raggedright\arraybackslash}p{4.0cm}
  >{\raggedright\arraybackslash}p{3.8cm}
}
\toprule
\multicolumn{4}{p{15.2cm}}{\textbf{
}\textit{Vision: A future grounded in collective well-being and shared prosperity, where AI is developed and governed as a shared and open societal resource in service of social and economic justice, personal agency, democratic participation, and environmental sustainability}.
} \\
\midrule

\textbf{Dimensions} & \textbf{Actions} & \textbf{Research priorities} & \textbf{Analytic considerations (emergent tensions)*} \\
\midrule

Openness definitions and gradients
& Mandate gradients of openness, including full or conditional openness of various models and datasets
& Define measures of openness, tradeoffs, and perceptions across stakeholders
& Differing views on whether openness should be mandatory, conditional, or treated as an end in itself \\

\midrule

Evaluation and standards
& Advance standards, evaluation practices, interoperability, and environmental safeguards
& Develop benchmarks, open evaluation ecosystems, and frameworks for safety and accountability
& \\

\midrule

Responsibility, safety, and usefulness
& Strengthen responsibility (e.g., safety and security practices, transparency, fairness) and ensure usefulness
& Assess safety frameworks, guardrails, and methods to capture social harms and misuse
& Tension between prioritizing expansion of openness and ensuring meaningful access and use for all \\

\midrule

Governance and institutions
& Establish participatory governance, embed data privacy and rights, and strengthen oversight
& Study effective governance models, roles of public and private actors, and regulatory approaches
& Unresolved questions around accountability, liability, and community governance \\

\midrule

Inclusive design, capacity building and communities
& Enable inclusive design; create platforms and literacy programs to broaden participation and fair recognition
& Explore training, labor models, and interdisciplinary collaborations that support inclusive design and diverse contributors
& \\

\midrule

Infrastructure and resources
& Build public infrastructure and ‘public option’ AI systems; supportive ecosystem for many models
& Investigate models for public compute, open datasets, and resource-sharing across regions
& \\

\midrule

Funding, economics, and incentives
& Develop sustainable funding and fair attribution systems; enabling and supporting developers of open source
& Design public goods funding, contributor compensation, and incentive structures
& \\

\midrule

Applications and public good
& Open models serve the public good and can be used in various applications, including smaller or purpose-built; considerations to social and environmental issues
& Assess impacts of open models as digital public goods in local, multilingual, and decentralized contexts; identify lessons from past technologies in relation to social and environmental impact
& \\

\bottomrule
\end{tabular}
\vspace{4pt}
{\footnotesize *Analytic considerations summarize cross-cutting tensions surfaced during synthesis rather than points of consensus.}
\end{table*}

\textit{Openness definitions and gradients.} Participants framed openness as operating along gradients, with different implications across data, models, software, and infrastructure. Actions included calls for open options across the AI stack, legal mandates compelling corporate models to be open (or at minimum publicly available for research and accountability purposes), and limits or conditions where full openness could produce harm---exposing a tension regarding whether openness should function as a compulsory baseline or be conditional. Research priorities focused on clarifying how openness can be defined and measured in practice, including developing frameworks to characterize and measure gradients of openness across the AI stack, as well as examining how different stakeholders (e.g., communities, companies and governments) understand and value openness. Additional priorities include mapping interlinkages between OSS, open data, open models, and open science communities; and studying how operational practices (e.g., governance, contribution management) shape meaningful openness beyond formal licensing.

\textit{Evaluation and standards.} Participants emphasized that openness without robust evaluation risks becoming symbolic rather than accountable. Evaluation and standards were framed as essential to making open AI systems reproducible, trustworthy, and legible. Action pathways included advancing shared standards for model and dataset release, embedding environmental sustainability criteria in funding and procurement processes, and strengthening transparency and consent norms in dataset sharing. Research priorities centered on developing rigorous, socially relevant evaluation frameworks, including domain- and task-specific benchmarks across the AI stack, open evaluation ecosystems, and testing regimes that extend beyond technical performance to capture safety, accountability, and societal impact.

\textit{Responsibility, safety, and usefulness. }Responsibility was framed as preventing harm and ensuring open AI systems are safe, usable, and meaningful in practice. Participants emphasized that openness alone does not guarantee responsibility. Action pathways focused on strengthening safety and security practices for open models, including safeguards against misuse, extractive deployment, and harmful downstream applications; while also prioritizing fairness, transparency, and interpretability. 
Several breakout groups included calls to prioritize smaller, more sustainable models over endless scaling. Research priorities included identifying safety, security, and risk-mitigation approaches specific to open models; clarifying which properties (e.g., explainability, privacy) can realistically be guaranteed in open settings; and developing safeguards against harmful and extractive use. Key research topics for usability include how to make open models more widely deployable and accessible to different people (e.g., locally runnable, multilingual, disability-inclusive, adaptable), while improving transparency and explainability for public understanding. A recurring tension concerned balancing the expansion of openness with the need to ensure meaningful access and safe use across diverse contexts.

\textit{Governance and institutions. }Governance discussions centered on who has authority and responsibility over open AI systems, how decisions are made and enforced, and how power is distributed. Participants emphasized moving beyond tokenistic consultation toward governance arrangements that grant communities real decision-making power. Action pathways included establishing participatory, multistakeholder governance structures for open AI models; promoting data privacy through federated approaches to data access and non-predatory data ecosystems that strengthen individual and community data rights; enabling model unlearning or withdrawal; and strengthening oversight through procurement, liability, and accountability mechanisms. Public procurement was identified as a key lever, with proposals to require open data standards, model documentation, and environmental disclosures, making openness the norm for publicly-funded systems while allowing for narrow exceptions when openness could endanger vulnerable groups. Research priorities focused on studying effective governance models that balance openness with accountability across public, private, and community actors. Key areas included collective and participatory governance approaches in open AI systems, comparative analyses of regulatory and legal frameworks, and empirical studies of decision-making around openness within governments, companies and large-scale open AI projects. Tensions emerged around accountability, liability, and how community governance, including context-dependent openness, can coexist (or not) with open source norms. 

\textit{Inclusive design, capacity building and communities.} Participants highlighted that openness is insufficient if communities lack the capacity to meaningfully participate in design, development, and use. Inclusive design and capacity building were framed as conditions for responsible openness rather than a downstream benefit, focusing on who is able to participate in design and development of open AI systems and under what conditions. Action pathways included broadening participation beyond a small number of geopolitical and institutional actors by investing in AI literacy and education, shared technical and institutional infrastructure, and community-led model development. Research priorities included examining AI training and education models, labor and recognition structures in open source development, and interdisciplinary collaborations that support sustained participation. Participants stressed that without attention to capacity and labor, open AI risks reproducing existing inequalities even when artifacts are publicly available.

\textit{Infrastructure and resources. } Participants emphasized that access to compute, tooling, and shared resources shapes who can build, evaluate, and contest open AI systems. Proposed actions included building public or shared compute infrastructure, supporting “public option” AI systems, creating experimentation sandboxes for open models and datasets, and developing shared hardware and tooling to reduce dependence on dominant infrastructure providers. Research priorities focused on examining infrastructural conditions under which openness in AI can be equitable and sustainable, including public compute infrastructure, cross-government data-sharing, and technical approaches to reducing infrastructure demands (e.g., distillation, quantization).

\textit{Funding, economics, and incentives. }Participants stressed that responsible openness requires sustainable economic conditions. Without long-term funding and fair attribution, open AI ecosystems risk fragility and capture. Action pathways included developing stable public and pooled funding mechanisms, compensating contributors, and creating incentives that reward safe, responsible, and environmentally conscious development. Research priorities focused on comparing funding models for open AI, assessing economic viability over time, studying incentive structures for participation and maintenance, and examining how costs and benefits are distributed across stakeholders. Illustrative cases of public investment in open source infrastructure were identified as valuable empirical sites for understanding how policy can support digital commons beyond large foundation models. 

\textit{Applications and public good. }Participants emphasized that openness should be considered and evaluated through its application and impact, not solely through artifact availability. Action pathways focused on supporting open AI models and datasets for application in civic technology, climate and environmental modeling, healthcare, scientific research, and public service delivery, often calling for and favoring smaller or purpose-built models. Research priorities included comparative studies of open versus closed approaches in different domains and contexts, assessments and historical analyses of past technologies to understand how openness, governance, and public investment have influenced shared prosperity. Environmental and social impacts were highlighted as cross-cutting concerns and areas for action and research, particularly as openness can increase reuse and deployment. 

\subsection{Tensions}
Despite general convergence around shared visions and a broad agenda for action and research, alignment coexisted with and illuminated unresolved complexity. Four core tensions became visible, particularly as participants translated aspirations into concrete actions, reflecting competing interpretations of how openness and open source AI should be enabled and mobilized in practice. These tensions emerged less as explicit disagreement and more as configurations that could not easily coexist, or in a few cases, as contradictions. They appeared most clearly in the action planning session, but surfaced throughout including in group discussions and share-outs.

\textit{Tension 1: Openness as an end vs. openness as a means.} Some contributions treated openness itself as the desired end state---as a value or principle to maximize in its own right. Others framed openness as instrumentally valuable when it supports broader social aims (e.g., equity, community benefit, shifting power hierarchies). This tension is linked to expectations placed on open source. In some cases, participants discussed how open source models and data can embody additional values and attributes, such as being ethically superior to closed models. This is reflected in one group’s actions that highlighted positive aspects of openness such as being more auditable and transparent, while also adding “open = inclusive… open = unbiased.” This perspective is tied to views that openness can inherently advance broader societal aims (e.g., inclusion) more effectively than closed models, and that ethical development can serve as a differentiator from closed models. As one participant noted, “attention on societal questions that aren’t just technical… may be better as an alternative approach [to corporate, closed models].” Others cautioned against or included explicit actions to \textit{not} have higher standards for open source models and datasets, as heightened expectations could create new burdens, disincentivize contribution, or slow development. One action emphasized, “Not imposing higher expectations (ethical, responsible) on open than [on] closed models.” This tension intersected with discussions of legal and accountability structures. Some inputs emphasized liability frameworks and accountability mechanisms for harms associated with open systems (aligned with openness as a means), while others called for legal protections or safe harbors to ensure that developers were not potentially penalized or faced repercussions for releasing open models (aligned with openness as an end). This tension is reflected in two seemingly contradictory actions, “Safe harbor for open developers” versus “Liability regime that holds developers accountable for foreseeable harms”. 

\textit{Tension 2: Expanding openness vs. enabling meaningful access.} Another tension concerned whether openness in itself is sufficient. This centers less on the purpose of openness and more on whether openness alone translates to real utility and access. In this, one orientation imagined openness as primarily a production task (e.g., creating and releasing more open models and datasets). Another emphasized that responsible openness requires meaningful access, usability across diverse contexts, and forms of participation and engagement. This tension highlights differing expectations about whether responsible openness is primarily about making things open or making openness work for different people. While these two orientations are not inherently incompatible, they reflect different priorities and resource commitments. Participants emphasized that genuine access requires education and support structures that enable different communities to meaningfully engage. Relatedly, some emphasized a “small is beautiful” approach, valuing purpose-built, contextually appropriate open systems that are driven by particular purpose and use cases, over large-scale openness.

\textit{Tension 3: Openness as mandatory vs. openness as conditional.} While there was broad alignment that openness is key to redistributing power away from dominant actors, contradictions emerged around how much openness is desirable and when it should be constrained (if at all). Some participants and breakout groups shared actions whereby openness and open-source is considered mandatory. Meanwhile, others emphasized model openness as a spectrum, with responsible openness requiring selective limitation or conditions. Actions developed in one group capture the contradiction explicitly: “Openness is not voluntary; it is compulsory,” alongside “Closed when necessary to protect socially marginalized communities.” This tension extended to datasets, with some actions calling for open data to support reproducibility and inclusion (e.g., “open data from the global majority [that could facilitate models that work for more people]”), and others highlighting how openness can reproduce extractive dynamics. Linked to these concerns were strong themes of consent, privacy, and protection, including recognition that some data, knowledge, or systems should not be fully open due to cultural, ethical, or safety considerations.

\textit{Tension 4: Openness itself is sufficient vs. value of openness is dependent on how it’s built, governed, and used.} A final tension concerned whether openness is sufficient on its own or whether its value depends fundamentally on how open data and models are developed, governed, and used---and by whom. Across the corpus, the idea of responsible openness was repeatedly conditioned on who is able to participate in building those systems, shaping decisions, and exercising power in open ecosystems. Several participants emphasized that openness does not automatically produce inclusive development or governance. One action called for “diverse participants in development/deployment [globally] to develop open source AI models that are responsive to their needs and interests,” emphasizing participation, stewardship, and community governance as central for responsible openness. At the same time, some contributions articulated a broader commons-oriented view, treating openness as the primary mechanism through which collective governance and inclusion could emerge. While these ideas are not conceptually incompatible, they reflect different assumptions about whether openness alone is sufficient to redistribute power. These unresolved questions surfaced in workshop discussions about building open foundation model ecosystems that are powerful enough to rival closed AI labs and genuinely decentralized, while not replicating the extractive dynamics they seek to counter.

\section{Discussion}
\label{sec:discussion}
\subsection{Reframing openness}
The workshop findings underscore a growing shift in how openness in AI is understood, away from binary choice between “open” and “closed”. Participants articulated openness as a spectrum of practices shaped by power, values, and social context. This perspective aligns with recent work arguing that openness in AI is relational and structured by institutional arrangements, governance choices and constraints \cite{Paris2025OpenAI} \cite{widder2024nature}. Some participants emphasized full transparency to advance accountability and innovation, while other participants called for conditional openness of models to support community needs for sovereignty or safety. This variability was not treated as a problem to be resolved, but a persistent tension requiring ongoing negotiation. Participants highlighted limits of treating openness as primarily artifact release, a framing that is more common in traditional OSS, and emphasized participatory structures that redistribute decision-making alongside technical openness. 

Moreover, openness was understood as extending beyond models to encompass data, compute infrastructure, governance structures, and other frameworks. This echoes \cite{Paris2025OpenAI} arguing openness should be understood at the level of AI ecosystems as opposed to individual models, and \cite{widder2024nature} noting that “open” narratives often obscure who can meaningfully participate. In our findings, openness as the level of one layer (i.e. model weights) was frequently seen as insufficient for meaningful access when other layers (e.g., data, compute, additional infrastructure) remain closed or limited. Relatedly, participants linked meaningful openness not only to technical access, but capacity. Several breakout groups emphasized genuinely open systems require appropriate AI literacy and support to engage with them in practice. This positions openness as inseparable from inclusion and capability, reinforcing concerns in prior work that openness without attention to use conditions can reproduce existing inequalities \cite{widder2024nature}. Taken together, findings show that openness in AI is not a neutral technical property, but a negotiated governance choice. Where systems land within the spectrum of openness, who decides the placement, and who has the capacity to utilize those systems fundamentally shapes whether openness redistributes or reinforces power.  

\subsection{Plurality of imaginaries}
The workshop revealed that a single, unified responsible open source AI future was neither clearly articulated across breakout groups nor necessarily desirable. While the workshop was initially designed with the expectation of convergence, participants instead generated visions shaped by their disciplinary, professional, and personal perspectives and values. Through synthesis, these diverse imaginaries coalesced around shared themes, producing coherence without uniformity. Importantly, participants consistently rejected the idea that responsible openness in AI corresponds to a single technological pathway or institutional solution. Instead of envisioning one definitive form of open source AI that is responsible, participants described multiple configurations of openness shaped by context, purpose, and community need. These plural imaginaries reflect different situated knowledges and informed the action pathways and research priorities. It may be tempting, particularly in response to concentrated power, to seek a singular solution, such as a publicly governed foundation model to serve as a universal alternative. The absence of a singular, definitive solution is not a weakness in the findings, but an empirical result reflecting value conflicts in responsible openness that cannot be resolved via technical design. This aligns with deliberative and participatory scholarship that critiques consensus-oriented approaches for suppressing conflict and minority perspectives \cite{mansbridge2006, mouffe2005} and emphasizes the value of “opening up” problem framings across diverse actors \cite{stirling2008}. 

Our contribution is empirical, showing that plurality need not prevent shared understanding, but can produce shared orientations that can guide action and research without premature closure. 
More broadly, findings reveal that confronting concentration of agenda-setting power in open AI requires ongoing, inclusive governance processes that can hold plural values; negotiating tradeoffs; and supporting multiple configurations of openness across contexts. 

\subsection{Tensions reflect values and positionalities}
The emergent tensions reveal that there are not clear “right” or “wrong” actions. Rather, the workshop exposed perspectives and positions rooted in different values, priorities, and positionalities. For some, openness itself was the primary ethical and political commitment: actions emphasized openness as an end in its own right, focusing on expanding openness and framing open source as mandatory---reflecting values tied to freedom, transparency, and distribution of power. Others held openness as meaningful when it translated into real capability and participation, prioritizing usability, meaningful access and use across communities---reflecting values of utility, inclusion, and practical benefit. And others surfaced openness as a means rather than an inherent good, emphasizing conditional openness, protections, and governance structures centering marginalized communities---reflecting values of equity and justice. In many cases these different perspectives and priorities could converge, whereas in other cases they resulted in contradictions. These different emphases and commitments help explain why definitions of “open source” AI itself, and what should legitimately count as sufficiently "open", become so contested. Prior work in STS and AI governance shows that standards and definitions are not neutral descriptors, but structuring devices that shape priorities, legitimacy, and resource allocation \cite{bowker1999, stirling2008, Paris2025OpenAI}. Disagreements about aspects such as whether openness should be compulsory or conditional are not merely semantic, but consequential for governance and accountability. 

Importantly, tensions require ongoing discussion, negotiation, contestation, and governance to inform how openness is pursued, supported, or constrained. Rather than ignoring tensions or treating divergence as a barrier to progress, plurality can be a productive condition: surfacing underlying values, inviting deliberation, and informing paths forward. Tensions also reveal that debates around open source AI cannot be reduced to a simple corporate versus non-corporate divide, but instead reflect deeper differences about values and purposes for openness to serve. Still, corporations can and do benefit from these unresolved tensions. While prior research has illustrated that companies embrace “open” AI and use rhetoric of “open” to expand market power \cite{widder2024nature}, our findings reveal that the openness spectrum itself affords companies a unique form of structural defensibility because legitimate value frameworks exist to justify varying degrees of openness. This inherent flexibility reinforces their power to define what counts as "responsible" openness and further emphasizes that a central challenge is not in identifying a single path to responsible openness in AI, but making plurality explicit and governable.  

\subsection{Methodological reflections}
The participatory process surfaced several methodological tensions that are informative for futures and participatory design research. A recurring challenge was ensuring shared understanding across participants with varying technical expertise. Creating pre-read materials was critical. At times, groups had to balance depth of discussion with inclusivity, occasionally limiting technical specificity to maintain collective engagement. This introduced tension, but also enabled cross-disciplinary thinking and generated higher-level reflections on values and governance. Relatedly, differences in technical expertise created moments of frustration about how far particular directions could be pursued. Rather than seeking resolution, the workshop sought to treat these differences as sites of negotiation and embrace plurality. 

\section{Conclusion}
\label{sec:conclusion}
As generative AI continues to permeate social, economic, and public life, questions of openness, governance, and accountability are urgent. This urgency is compounded by significant public investment alongside concentrated decision-making power over definitions and pursuit of "openness". This paper used futures thinking and participatory design methods to examine how diverse stakeholders could collectively imagine and negotiate responsible openness in AI. Our workshop surfaced co-created visions, action pathways, and research priorities---alongside productive tensions revealing that openness is an ongoing sociotechnical project informed by values, priorities, and positionalities. 
While our workshop included global representation, a key limitation is that workshop participants were primarily US-based and did not fully reflect the breadth of the global open source and AI ethics communities. Further, findings may be influenced by selection bias, though we sought to mitigate this through the snowball sampling and recommendation approach.
Additionally, while participatory design methods surface collective and situated insights that individual-level methods cannot replicate, they are not designed for statistical generalizability.
Future work should recruit more globally diverse participants, particularly from low- and middle-income countries, and complementary methods such as large-scale surveys can extend this work by examining how tensions and priorities resonate across broader communities of practice. Future work could also examine how the tensions surfaced here play out across the field broadly, including which value frameworks come to dominate, through what mechanisms, and whose perspectives shape versus get sidelined in ongoing debates about what responsible openness looks like in practice. 
Methodologically, this paper demonstrates the role participatory futures and design approaches can play in AI governance and agenda setting.  
This work argues for AI governance approaches that embrace plurality, make tensions visible, and clarify how decisions about openness are made, contested, and managed. 

\section*{Generative AI Usage Statement}
Generative AI (Gemini) was used to assist with grammar, writing fluency, and table formatting.

\section*{Author Contributions}
\textbf{Genevieve Smith}: Conceptualization, Methodology, Investigation, Formal Analysis, Writing – Original Draft, Writing – Review \& Editing, Supervision, Project Administration.
\textbf{Hiral Patel and Steven Luo}: Investigation, Data Curation, Writing – Original Draft, Writing – Review \& Editing. \textbf{Shachee Doshi, Nicholas Garcia, Woohyeuk Lee, Jarrod Millman, Cailean Osborne, Katie Steen-James, Nikko Stevens, Jennifer Tridgell, David Gray Widder}: Writing – Original Draft, Writing – Review \& Editing.
\textbf{Monica Bobra, Judy Brewer, Cathryn Carson, Isadora Cruxen, Maximilian Gahntz, Meredith Lee, Min Kyung Lee, Natalia Luka, Chinasa Okolo, Ricardo Mirron Torres, Derek Slater}: Writing – Review \& Editing.

\begin{acks}
The authors would like to acknowledge all participants who gave their time and expertise to the workshop on responsible openness and open source in AI. The workshop was supported by the National Science Foundation (Award ID 2427618), with additional support from Mozilla. The views and findings expressed in this paper are those of the authors and do not necessarily reflect the views of the National Science Foundation or Mozilla.
\end{acks}

\bibliography{acmart}

\clearpage
\begin{appendices}
\pagestyle{headings} 
\section{Terms: Degrees of Openness in AI Systems}
\label{sec:appendix}

The categories below reflect commonly referenced terms in current technical and policy debates. They are used descriptively to clarify technical practices. The categories are presented in approximate order: systems in higher rows meet all requirements of lower rows, but not vice versa. This terminology is used to disambiguate how “open source” and “open” are commonly invoked, rather than to resolve normative debates about what openness should entail.

\renewcommand{\arraystretch}{1.2}

\begin{table}[htbp]
\centering
\caption{Degrees of Openness in AI Systems}
\label{tab:ai-openness}
\begin{tabular}{|p{3.5cm}|p{7cm}|p{4cm}|}
\hline
\textbf{Term} & \textbf{About} & \textbf{Examples} \\
\hline
\textbf{Fully open source AI} &
Systems releasing complete source components, including training data, model architecture, training code, weights, and documentation; everything needed for independent reproduction and modification. &
Ai2’s OLMo and Tülu~3 models (full pipeline); MIT’s Boltz-1 \\
\hline
\textbf{OSAID-compliant AI} &
Systems meeting the Open Source Initiative’s Version~1.0 definition: open weights, architecture, inference code, and \emph{“sufficiently detailed data information,”} but not necessarily complete training data. &
EleutherAI’s Pythia; LLM360’s Amber and CrystalCoder; Google’s T5 \\
\hline
\textbf{Open weights models} &
Systems releasing trained model weights and inference code while withholding training data. &
Meta’s LLaMA-2; DeepSeek-R1; Mistral models \\
\hline
\textbf{Partially open models} &
Systems exhibiting selective transparency: some components publicly available under open licenses, while others remain restricted. &
OpenAI’s GPT-OSS (some weights released under Apache~2.0) \\
\hline
\end{tabular}

\vspace{0.5em}
\small
\textit{Note.} This treats “fully open source AI” as requiring the same completeness as traditional open source software, not as synonymous with OSAID compliance. The OSAID Version~1.0 acceptance of “data information” rather than training data remains contested.
\end{table}

\end{appendices}

\end{document}